\documentclass[amsmath,amssymb,superscriptaddress,aps,prl,twocolumn,reprint]{revtex4-1}

\usepackage{graphicx}
\usepackage{subfig}

\usepackage{amssymb,amsfonts,amsmath}
\usepackage{bm}
\usepackage{color}

\renewcommand {\vec} [1] {{\bm #1}}
\newcommand{\ff}{g}
\newcommand{\tf}{h}
\newcommand{\mat}[1]{#1}
\newcommand{\noise}{\eta}

\begin{document}


\title{Application of compressed sensing to the simulation of atomic
  systems}


\author{X. Andrade}
\author{J. N. Sanders}
\author{A. Aspuru-Guzik}
\email{aspuru@chemistry.harvard.edu}
\affiliation{Department of Chemistry and Chemical Biology, Harvard
  University, 12 Oxford Street, Cambridge, MA 02138, United States}




\begin{abstract}
  Compressed sensing is a method that allows a
  significant reduction in the number of samples required for accurate
  measurements in many applications in experimental sciences and
  engineering. In this work, we show that compressed sensing
  can also be used to speed up numerical simulations. We
  apply compressed sensing to extract information from the real-time
  simulation of atomic and molecular systems, including electronic and nuclear
  dynamics. We find that for the calculation of vibrational and
  optical spectra the total propagation time, and hence the
  computational cost, can be reduced by approximately a factor of
  five.
\end{abstract}

\maketitle






\section{Introduction}

A recent development in the field of data analysis is the compressed
sensing (CS) (or compressive sampling)
method~\cite{Candes2006,Donoho2006}. The foundation of the method is
the concept of \emph{sparsity}: a signal expanded in a certain basis
is said to be sparse when most of the expansion coefficients are zero.
This extra information can be used by the CS method to significantly
reduce the number of measurements needed to reconstruct a signal. CS
has been successfully applied to data acquisition in many different
areas~\cite{Baraniuk2010}. For example, to improve the resolution of
medical magnetic-resonance imaging~\cite{Lustig2007}. It has also been
applied to the experimental study of atomic and quantum
systems~\cite{Shabani2011,AlQuraishi2011,Katz2010}.


In this article we show that CS can also be an invaluable tool for
some numerical simulations where the optimal sampling of CS is
reflected in a considerable reduction of the computational cost. We
focus on atomistic simulations of nanoscopic systems by using CS to
extract frequency-resolved information from real-time methods such as
molecular dynamics (MD) and real-time electron dynamics. 

In MD\cite{Rapaport1995,Allen1989} the trajectory of the atomic nuclei
is obtained by integrating their equations of motion with an
interaction obtained from parametrized force-fields or by explicitly
modeling the electrons~\cite{Marx2009}. Many static and dynamical
properties can be obtained from MD, making it one of the most widely
used methods to study atomistic systems computationally. As such it is
important to develop methods that can improve the precision and reduce
the computational cost of this method, especially for ab-initio
MD.

While not as widely used as MD, real-time electron dynamics, in
particular real-time time-dependent density functional theory
(TDDFT)~\cite{Runge1984}, is an important approach to study linear and
non-linear electronic response~\cite{Yabana1996,Castro2004c,Castro2004b,Takimoto2007}. Due to the scalability and
parallelizability properties real-time TDDFT is particularly efficient
for large electronic systems~\cite{Andrade2012}, so an additional
reduction in the computational cost can push the boundaries of the
system-sizes that can be studied.

Many physical properties are represented by frequency-dependent quantities. To
obtain these from a real-time framework usually a discrete
Fourier transform (FT) is used. Our approach is to replace this FT by
a calculation of the Fourier coefficients based on the CS method. To
obtain a given frequency resolution, the CS method requires a total
propagation time that is several times smaller than that required by a
FT.

CS has the potential to provide \emph{across-the-board} speedup for
many applications involving the computation of sparse spectra.
Moreover, this speedup may be obtained without making \emph{any}
changes to the underlying propagation code used in different types of
electronic and nuclear calculations; one simply replaces the FT
algorithm with the CS method, making the approach quite
straightforward to implement.  This paper introduces CS and
demonstrates its broad utility in computational chemistry and physics
by applying it to the calculation of various nuclear and electronic
spectra of small molecules. The resulting computer code is available
as open-source software.

The article is structured as follows. We first introduce the CS method
and show how it may be applied to the determination of Fourier
coefficients.  Next, we apply CS to the calculation
of vibrational, optical absorption, and circular dichroism spectra.
We then proceed to a discussion of the numerical methods used in our
CS implementation. Finally, we offer conclusions and an outlook.

\section{Compressed sensing}

In this section, we briefly introduce the application of the
CS method to the calculation of Fourier coefficients. More details
about the method and its origins may be found in
Refs.~\cite{Baraniuk2007,Candes2008,Chartrand2010}.

For simplicity, we assume that we want to calculate a certain
frequency-resolved quantity \(\ff(\omega)\) that is given by the \emph{sine}
transform of a certain time-resolved function \(\tf(t)\)
\begin{equation}
  \label{eq:sigmai}
  \ff(\omega) = \int_{-\infty}^\infty dt \, \sin({\omega t})\, \tf(t)
\end{equation}
(the analysis is equally valid for the cosine transform).
Since we are interested in numerical solutions, we need to think in
terms of discrete quantities. In this case we want to obtain a series
of values \(\{\ff_1, \ff_2, \ldots, \ff_{N_\omega}\}\) at \(N_\omega\)
equidistant frequencies \(\omega_j=\Delta\omega\,j\), from the known
set of time-resolved values $\{\tf_1, \tf_2, \ldots, \tf_{N_t}\}$ given at \(N_t\) equidistant times \(t_j=\Delta t j\).

In principle, the \(\ff_k\) set can be directly obtained using the
discrete FT,
\begin{equation}
  \label{eq:ft}
  \ff_k = \sum_{j=1}^{N_t} \Delta t\sin(\omega_k t_j)\,\tf_j\ .
\end{equation}
However, if we expect that many of the Fourier coefficients are zero,
a property known as \emph{sparsity}, we can use this additional
information to obtain more precise results. This is the basis for the
CS scheme.

We start by reformulating the discrete Fourier transform in
eq.~(\ref{eq:ft}) as a matrix inversion problem. From this
perspective, we are trying to solve the linear equation for
~\(\vec{\ff}\),
\begin{equation}
  \label{eq:linear}
  \mat{F}\vec{\ff}=\vec{\tf}\ ,
\end{equation}
where \(\mat{F}\) is the \(N_\omega\times N_t\) Fourier matrix with entries
\begin{equation}
  \label{eq:f}
  F_{jk} = \frac{2}{\pi}\Delta\omega\sin(\omega_j t_k)\ .
\end{equation}

Our objective is to obtain sensible results with \(N_t\) as small as
possible. Thus, we are interested in the case \(N_\omega > N_t\), where
the linear system is under-determined, and there are many solutions for
\(\vec{\ff}\) (in fact, one of them will be given by
eq.~(\ref{eq:ft})). From all the solutions of eq.~(\ref{eq:linear}),
we select the one that has the largest number of zero coefficients:
the sparsest solution. This turns out to be equivalent to
the so-called basis-pursuit (BP) optimization problem~\cite{Candes2008}
\begin{equation}
  \label{eq:bp}
  \min_{\vec{g}} |\vec{\ff}|_1 \quad \textrm{subject to}\quad \mat{F}\vec{\ff}=\vec{\tf}\ ,
\end{equation}
which is what one solves in practice (where \(|\ff|_1=\sum_k|\ff_k|\) is the standard 1-norm).


The CS scheme can be generalized to allow for a certain amount of
noise in the time-resolved signal. In this case the problem to be
solved is known as basis-pursuit de-noising (BPDN)
\begin{equation}
  \label{eq:bpdn}
  \min_{\vec{g}} |\vec{\ff}|_1 \quad \textrm{subject to}\quad
  \left|\mat{F}\vec{\ff}-\vec{\tf}\right| < \noise\ ,
\end{equation}
where \(\noise\) represents the level of noise in the signal. This is
the formulation we use in our case, since we expect a certain
amount of noise coming from the finite-precision numerical
calculations (and possibly other sources).

\section{Vibrational spectra}


MD can be used to obtain information about the vibrational modes of
atomic systems. Experimentally, the quantities that usually give
access to the vibrational modes are the infrared and Raman spectra
that can be obtained from MD as the Fourier components of the
electronic polarization and polarizability, respectively. If we are
only interested in the vibrational frequencies, from the nuclear
velocities, \(\left\{\vec{v}_i\right\}\), we can calculate the
velocity autocorrelation function
\begin{equation}
    \gamma(t) =\frac{\left<\sum_i\vec{v}_i(t)\cdot\vec{v_i}(0)\right>}{\left<\sum_i\vec{v}_i(0)\cdot\vec{v}_i(0)\right>}
    \ , \label{eq:vaf}
\end{equation}
whose cosine transform is the vibrational frequency distribution~\cite{Dickey1969}
\begin{equation}
    f(\omega) =\int\mathrm{d}t\,\gamma(t)\,cos(\omega t)\ . \label{eq:vibspect}\\
\end{equation}
Since this spectrum is composed of a finite number of frequencies
(less than three times the number of atoms in the system), the
calculation is ideal for the CS method.

\begin{figure}
  \centering
  \includegraphics*[width=\columnwidth]{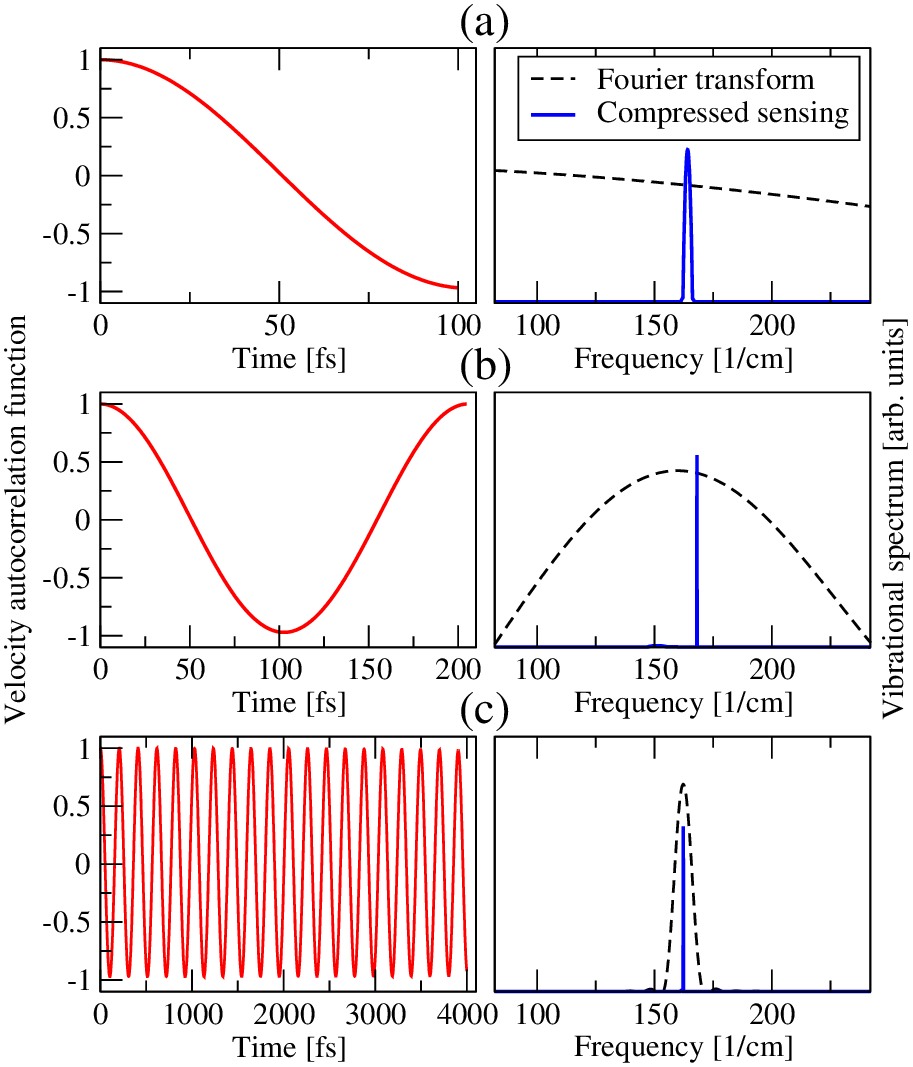}
  \caption{Frequency distribution spectrum of Na\(_2\) calculated
    using a Fourier transform and compressed sensing for different
    total propagation times: a) 100 fs, b) 205.65 fs (\(\approx 1\)
    oscillation period), and c) 4000 fs. The left plots show the
    velocity autocorrelation function and the right plots show the frequency spectrum.}
  \label{fig:diatomic}
\end{figure}

\begin{figure}
  \centering
  \includegraphics*[width=\columnwidth]{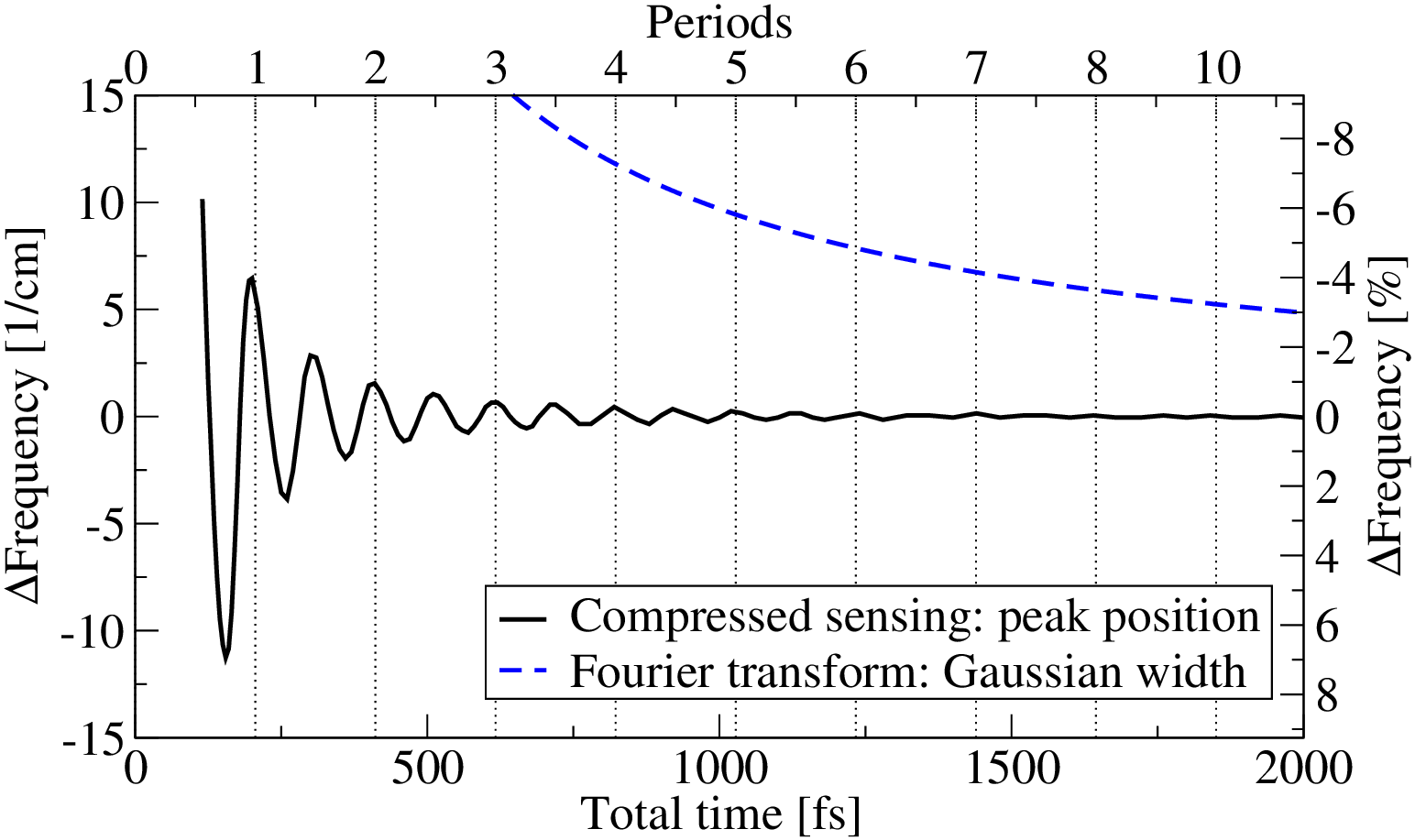}
  \caption{Error in the vibrational frequency of Na\(_2\) computed by
    compressed sensing with respect to total time. For comparison
    we plot the width of the peak obtained by using a discrete Fourier
    transform. The width, \(\sigma\), is calculated by assuming the
    peak has a Gaussian form \(A\exp[-\omega^2/(2\sigma^2)]\).}
  \label{fig:diatomic_conv}
\end{figure}

To illustrate the properties of the CS method we start with a simple
case, the single vibrational frequency of a diatomic molecule,
Na\(_2\), that we simulate using ab-initio molecular dynamics. In
Fig.~\ref{fig:diatomic}, we show how the vibrational spectrum depends
on the amount of time for which the velocity autocorrelation is
calculated. While the discrete FT requires long times to resolve the
vibrational frequency, the CS method gives a
well-defined peak even with less than one oscillation of the
molecular vibrational mode. That the peak is well defined, however, does not imply that
peak position is converged. As it can be seen in
Fig.~\ref{fig:diatomic_conv}, the peak position oscillates with the
total time until it converges to the proper value after a few periods
are sampled. Still, the result converges much faster than compared
with the width of the peak given by a FT. We remark that the CS
process does not use any additional information about the the signal
beyond assuming it is sparse.

\begin{figure}
  \centering
  \includegraphics*[width=\columnwidth]{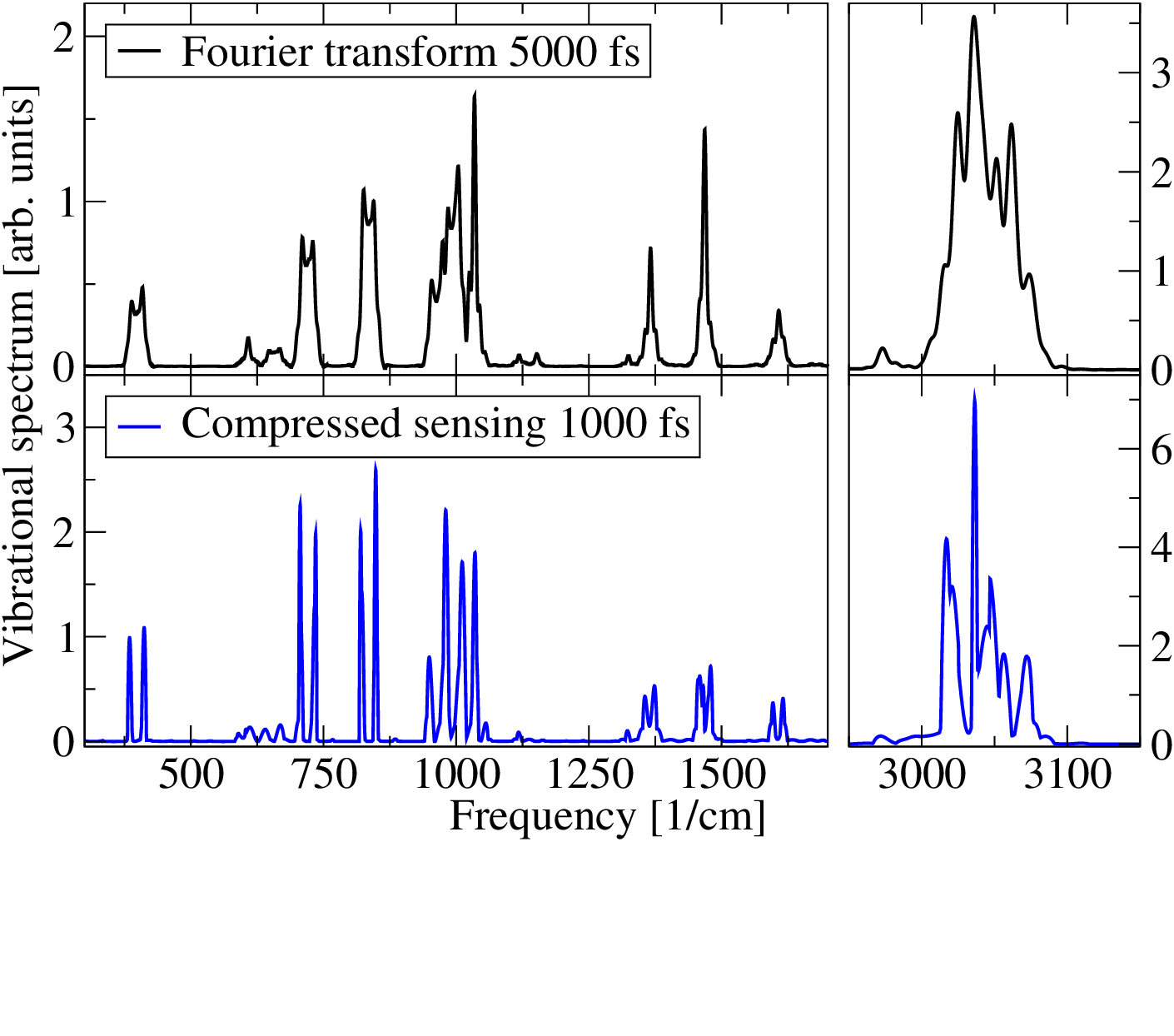}
   \caption{Frequency distribution spectrum of benzene. Comparison of
     a compressed sensing calculation with 1000 fs and a Fourier
     transform with 5000 fs.}
  \label{fig:vibrational}
\end{figure}

To further demonstrate the advantages of this approach, we now calculate the vibrational spectrum for a benzene molecule from
a ab-initio MD simulation, Fig.~\ref{fig:vibrational}. We can see that
the CS approach with 1000 fs obtains a spectrum that is
better resolved than the FT results for 5000 fs. This is directly translated into a
reduction of the computational time by five times or more. It is reasonable to
expect that equivalent gains can be obtained for the computer simulation of other vibrational
spectroscopies like infrared and Raman.

\section{Optical absorption spectra}

Optical absorption is an electronic process. While it can be
calculated from a linear response
framework~\cite{Casida1996,Andrade2007}, it can also be obtained from
real-time electron dynamics~\cite{Yabana1996}.
To obtain the spectrum from real-time
dynamics the electronic system is propagated under the effect
on an electric field of the form \(\vec{E}(\vec{r},t) =
\vec{\kappa}\delta(t)\). From the propagation the time-dependent
dipole moment \(\vec{p}(t)\) is obtained, and from the dipole, the
frequency-dependent polarizability can be obtained as (atomic units
are used in the next two sections)
\begin{equation}
 \alpha_{ij}(\omega) = \frac{1}{\kappa_i} \int_0^\infty dt \,
 e^{-i\omega t}\left[p_j(t) - p_j(0)\right]\ .
\end{equation}
In order to obtain the full \(\vec{\alpha}\) tensor three propagations are
required (with \(\vec{\kappa}\) in different directions.

The absorption cross-section is related to the trace of the imaginary part of the polarizability tensor
\begin{equation}
  \label{eq:absorption}
  \sigma(\omega) = \frac{4 \pi \omega}{3c}\mathfrak{Im}\sum_{i}\alpha_{ii}(\omega)\ .
\end{equation}
The optical absorption spectra is an ideal candidate for the application of CS.  For a molecule, the electronic transitions between bound states produce a discrete spectrum in the low energy region. At higher energies, the transitions to unbound states produce a continuous spectrum. Standard calculation approaches, however, cannot capture this continuous spectrum and approximate it as a sequence of discrete excitations.

\begin{figure}
  \centering
  \includegraphics*[width=\columnwidth]{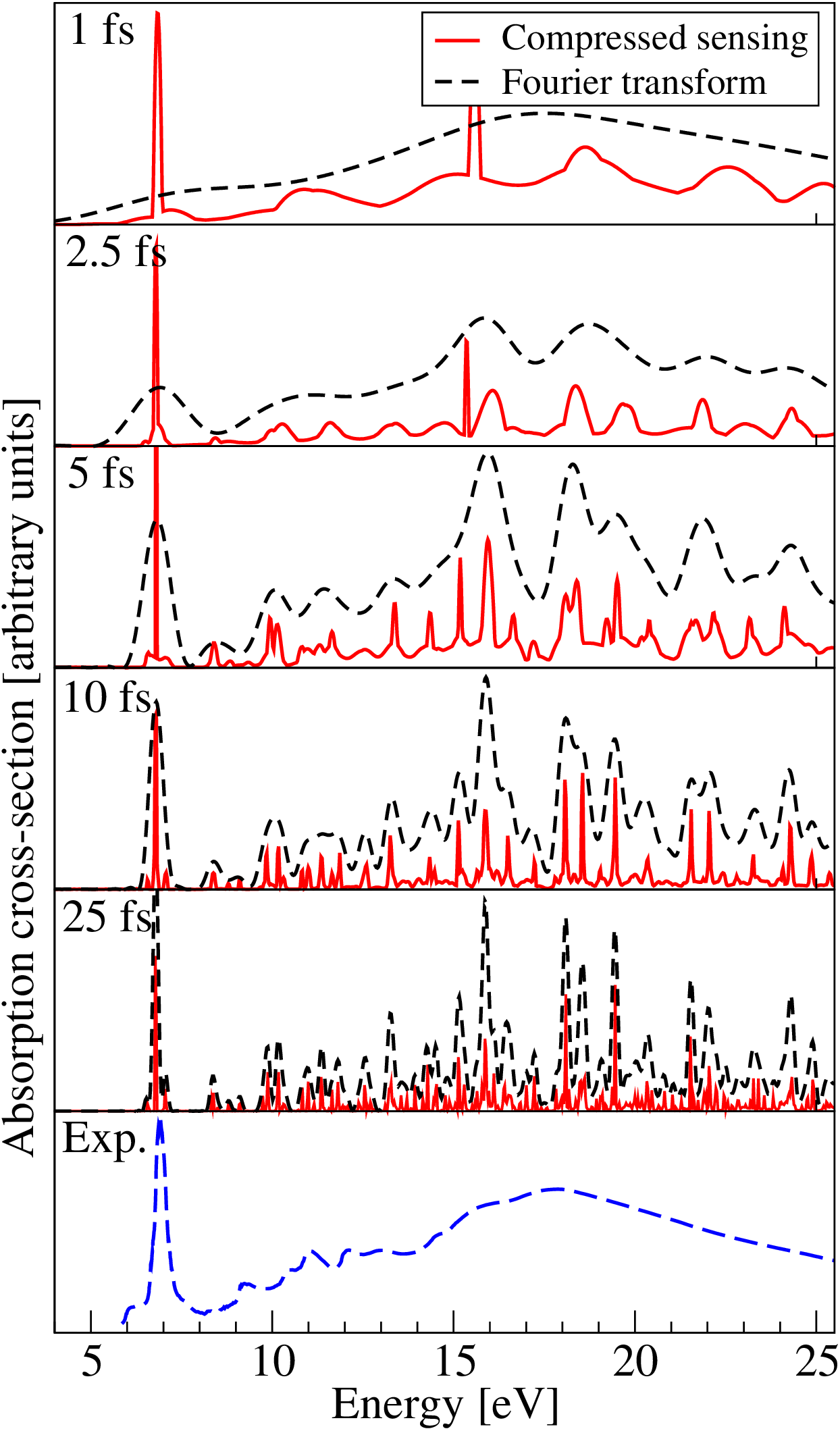}
  \caption{Optical absorption spectra for benzene computed from
    real-time TDDFT with different propagation times. Comparison between compressed sensing and discrete Fourier transform. Experimental
    results from Ref.~\cite{Koch1972CPL}.}
  \label{fig:benzene}
\end{figure}

In Fig.~\ref{fig:benzene}, we show the optical absorption spectrum for
benzene calculated via real-time TDDFT. There we illustrate the effect
of the propagation time on the spectrum for CS and FT. From the
figure, it is clear that the CS method is capable of resolving the
spectrum much better and with a shorter propagation time than a
discrete FT.  For a given resolution, the FT requires approximately 5
times the propagation time as CS (as can be seen, for example, by
comparing the FT at 25 fs with CS at 5 fs).

\section{Circular dichroism spectra}

Another property that can be calculated from real-time electron
dynamics is circular dichroism (CD)
spectra~\cite{Yabana1999,Varsano2009}.  A CD spectrum measures the
difference in a chiral molecule's response to left and right
circularly-polarized light.  The real-time calculation is performed in
the same manner as the optical absorption case, but the key quantity
to be calculated is now the time-dependent orbital magnetization
\(\vec{m}(t)\). The frequency-dependent electric-magnetic
cross-response tensor may be obtained as
\begin{equation}
  \label{eq:beta}
  \beta_{ij}(\omega) = \frac{ic}{\omega \kappa_i} \int_0^\infty dt \,
  e^{-i\omega t}\left[m_j(t)-m_j(0)\right] \ .
\end{equation}
The rotatory strength, which is the quantity typically plotted in CD
spectra, is related to the trace of the imaginary part of
\(\vec{\beta}(\omega)\) according to
\begin{equation}
  \label{eq:dichroism}
  R(\omega) = \frac{\omega}{\pi c}\mathfrak{Im}\sum_{i}\beta_{ii}(\omega)\ .
\end{equation}
The rotatory strength \(R(\omega)\) is suitable to the CS scheme
because it is sparse in frequency space.  In fact, the peaks in a CD
spectrum are located at the same positions as in an absorption
spectrum. However, the CD spectrum contains both positive and
negative peaks.

\begin{figure}
  \centering
  \includegraphics*[width=\columnwidth]{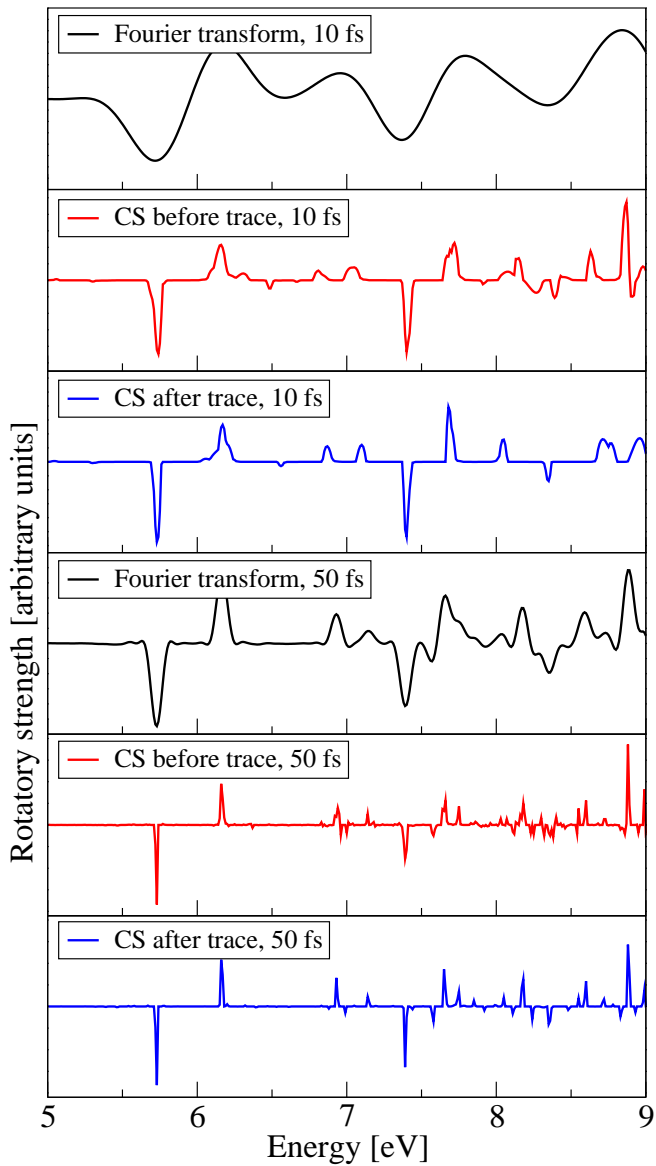}
  \caption{Circular dichroism spectrum computed for
    (\emph{R})-methyloxirane from real-time TDDFT with different
    propagation times. Comparison between discrete Fourier transform
    and compressed sensing (CS). Since the CS process is non-linear we
    compute the spectra in two ways. CS before trace: the spectra is
    calculated for each direction using CS and then the trace,
    eq.~(\ref{eq:dichroism}), is computed. CS after trace: the trace is
    calculated in the time domain and then the CS process is used over
    this averaged signal.}
  \label{fig:methyloxirane}
\end{figure}

Fig.~\ref{fig:methyloxirane} compares the CD spectrum for
(\emph{R})-methyloxirane as computed by FT and CS for two different
propagation times (10 fs and 50 fs).  As can be seen from the figure,
for a given propagation time, the CS method provides better spectral
resolution than the discrete FT.  In fact, just as with linear
absorption, FT requires a propagation time approximately 5 times as
long as CS to obtain a comparable spectral resolution (as can be seen
by comparing the 50 fs FT with the 10 fs CS).

Fig.~\ref{fig:methyloxirane} also illustrates another feature of CS:
unlike a direct FT, the CS method is non-linear.  Adding together
time-resolved signals, then applying CS, generally gives different
results from applying CS first and then adding together the results in
the frequency domain; this is particularly the case if not all the
peaks are well-resolved.  In other words, the use of CS to convert
time-resolved data into the frequency-domain, as in eq.~(\ref{eq:beta}),
and the calculation of the trace, as in eq.~(\ref{eq:dichroism}), do not
commute. Hence, there are two approaches to obtain the CD spectrum: we
can perform CS for each propagation direction and then compute the
trace, or we can compute the trace in the time-domain and then perform
CS. Both approaches are shown in Fig.~\ref{fig:methyloxirane}; at 10
fs, they give similar but not identical results.  This is to be
expected: the ability of CS to resolve peaks depends on the sparsity
of the spectrum. Since each propagation direction is sparser than the
sum of all three, CS resolves more peaks at 10 fs when it is performed
prior to the trace. For longer propagation times (50 fs), all of the
peaks are more fully resolved and the two approaches converge.  In any
event, both approaches to CS provide much improved resolution over a
direct FT for a given propagation time.

\section{Numerical methods}
\label{sec:methods}

Numerically, to find a spectrum using the CS method we need to solve
eq.~(\ref{eq:bpdn}). This is not a trivial problem, so we rely on the
SPGL1 algorithm developed by van~den~Berg and
Friedlander~\cite{Berg2008}. To avoid numerical stability issues we
work with a normalized BPDN problem, where the factors of the \(F\)
matrix, eq.~(\ref{eq:f}), are left out and \(\vec{\tf}\) is
normalized. The missing factors are included in \(\vec{\ff}\) after
the solution is found. This has the additional advantage of making the
noise parameter \(\noise\) of eq.~(\ref{eq:bpdn}) dimensionless.

Since we do not a have an {\it a priori} estimate for \(\noise\), we
do not set it directly. As the SPGL1 algorithm finds a sequence of
approximated solutions with a decreasing value of \(\noise\), we set
the target value to zero. We assume that the calculation is converged
when the value of \(\noise\) falls below a certain threshold
(\(10^{-7}\)) or the active space of the system, the set of non-zero
coefficients, has not changed for a certain number of iterations
(50). In the former case we consider that a solution of the BP
problem, eq.~(\ref{eq:bp}), has been found. For all the calculations
presented here \(\noise < 10^{-3}\).

CS is much more costly numerically than the discrete FT approach, as
it usually involves several hundreds of matrix
multiplications. However, this is not a problem for application since
usually the solution process normally only takes a few minutes, much
less than the computation time required to simulate the real-time
dynamics of large atomic systems.

All the calculations presented in this article were performed using
the {\sc octopus} code~\cite{Castro2006,Andrade2012} at the (time-dependent)
density functional theory level with the PBE exchange correlation
functional~\cite{Perdew1996}. The adiabatic molecular dynamics
calculations were performed from first principles using the modified
Ehrenfest method~\cite{Alonso2008,Andrade2009} with a \(\mu\) factor
of 30 for Na\(_2\) and 5 for benzene. The systems were given initial
velocities equivalent to 300 K and the MD is performed at constant
energy. 

All calculations used norm-conserving pseudo-potentials with a
real-space grid discretization. The shape of the grid is a union of
boxes around each atom. For Na\(_2\) we use a spacing of
0.375~a.u. with a sphere radius of 12~a.u., and the MD time-step is
0.057 fs. For benzene, the grid-spacing is 0.35~a.u., the radius is 14~a.u., and the time-step is 0.0085 fs for MD and 0.0017 fs
for real-time TDDFT. For (\emph{R})-methyloxirane, the spacing is
0.378~a.u., the sphere radius is 15.1~a.u., and the time-step is
0.0008 fs for real-time TDDFT.  For the vibrational spectrum
calculation we use a time-step 10 times the one of the MD, the energy
step is 0.01 1/cm, and the maximum spectrum energy is 5000 1/cm. For
the benzene optical absorption spectra, we use a time-step of
0.0017~fs, the energy step is 0.027~eV, and the maximum spectrum
energy is 820~eV.  For the (\emph{R})-methyloxirane circular dichroism
spectra, the time-step is 0.0008~fs, the energy step is 0.01~eV, and
the maximum spectrum energy is 330~eV. The structure of benzene was
taken from ref.~\cite{Curtiss1997} and the structure of
(\emph{R})-methyloxirane was taken from ref.~\cite{Carnell1991}. 

All discrete FTs were performed using third-order polynomial damping:
each signal at time \(t\) was multiplied by \(p(t) = 1 - 3(t/T)^2 +
2(t/T)^3\) prior to Fourier transform, where \(T\) is the time-length
of the signal.

The SPGL1 method used for CS was implemented into {\sc octopus} based
on a Fortran translation of the original Matlab code of van~den~Berg
and Friedlander~\cite{Berg2007}. We plan to release this
implementation as a standalone tool in the near future (for the moment
the code can be obtained from the {\sc octopus} repository).

\section{Conclusions}

We have shown that the CS method can be applied to the numerical
calculation of different kinds of atomic and electronic spectra. This
results in a significant reduction of the computational time required
for the numerical simulations. The effect of this reduction is to
increasing the size of the systems that are currently accessible to
numerical simulations, and to make possible simulations with more
precise, but more costly, methods. It also means that other types of
simulations become more affordable from a real-time perspective, for
example the combined dynamics of nuclei and electrons that are
constrained to short simulation times by the fast dynamics of the
electron.

In this work, we have shown the application of CS to the calculation of
a few types of spectra, but the method most likely can be applied to
other quantities as well, such as non-linear optical response~\cite{Castro2004b,Takimoto2007},
magnetic circular dichroism~\cite{Lee2011}, semi-classical nuclear
dynamics~\cite{Ceotto2009,Ceotto2011}, 2D
spectroscopy~\cite{Mancal2006, YuenZhou2011}, etc. Of course, the
method is not limited to atomistic simulations and could be applied to
simulations in all scientific fields.

The main limitation of the CS approach is that it will not be
beneficial for quantities that are not sparse. In such case, the
performance of CS will be equivalent to a standard discrete FT. There
are some cases where the sparsity requirement might be
circumvented. For example, though the real part of the polarizability
tensor is not sparse, it could be computed from the imaginary part by
using the Kramers-Kronig relation. Another example is crystalline
systems~\cite{Bertsch2000}, where there is a continuum of excitation energies. In this
case it might be possible to apply the CS scheme to each \(k\)-point
separately.

We expect that compressed sensing will become widely used in the
scientific computing community once its advantageous properties become
more widely known. The main difficulty in the adoption of CS is that it is more
complex to implement than a discrete FT. This problem can be solved by
providing libraries and utilities that can be used by researchers as a
black-box. Other issue of the CS approach it has some aspects that
might result counter-intuitive at first. For example, non-linearity
and the fact that with CS the peak width is not always related to
the convergence of the spectrum.

Moreover, we believe that our direct application of the compressed
sensing methodology to numerical simulation opens the path for more
challenging applications. An idea that we could call ``compressed
computing'', where the principles of sparsity could be used to design
algorithms for numerical simulations that have a reduced computational
cost of calculations not only in the number of operations, but also in
memory and data transfer bandwidth requirements.





\begin{acknowledgments}
  We acknowledge S.~Mostame, J.~Yuen-Zhou, M.-H.~Yung, J.~Epstein,
  M.\,A.\,L.~Marques, and A.~Rubio for useful discussions. The
  computations in this paper were run on the Odyssey cluster supported
  by the FAS Science Division Research Computing Group at Harvard
  University.

  This work was supported by the Defense Threat Reduction
Agency under Contract No HDTRA1-10-1-0046 and by the Defense Advanced Research Projects Agency under award number N66001-10-1-4060.  J.N.S. acknowledges support from the Department of Defense (DoD) through the National Defense Science \& Engineering Graduate Fellowship (NDSEG) Program.  Further, A.A.-G. is grateful for the support of the Camille and Henry Dreyfus Foundation and the Alfred P. Sloan Foundation.

\end{acknowledgments}






%
\bibliography{biblio}









\end{document}